\documentclass[twocolumn,prx,nobibnotes,altaffilletter,amsmath,amssymb,amsfonts]{revtex4-1}
\usepackage[latin1]{inputenc}
\usepackage{graphicx}
\usepackage{amsmath}
\usepackage{latexsym}
\usepackage{epsfig}
\usepackage{bm}	
\usepackage{enumerate}
\usepackage{lipsum}

\usepackage[dvipsnames]{color}

\begin{document}
\title{Velocity distribution in active particles systems}
%\title{Multidimensional Unified Colored Noise Approximation for \\
%Stationary Probability Distributions in 
%Active Matter}

\author{Umberto Marini Bettolo Marconi$^1$}
%\email{claudio.maggi@roma1.infn.it}
\author{Nicoletta Gnan$^2$}
\author{Matteo Paoluzzi$^3$}
\author{Claudio Maggi$^4$}
\author{Roberto Di Leonardo$^{4,5}$}

\affiliation{$^1$Scuola di Scienze e Tecnologie, Universit\`a di Camerino, Via Madonna delle Carceri, 62032, Camerino, INFN Perugia, Italy}
\affiliation{$^2$CNR-ISC, UOS Sapienza, P.le A. Moro 2, I-00185, Roma, Italy}
\affiliation{$^3$Physics Department, Syracuse University, Syracuse, NY 13244}
\affiliation{$^4$Dipartimento di Fisica, Universit\`a di Roma ``Sapienza'', 
I-00185, Roma, Italy }
\affiliation{$^5$NANOTEC-CNR Institute of Nanotechnology, Soft and Living Matter Laboratory Piazzale A. Moro 2, I-00185 Roma, Italy
}

\date{\today}
\begin{abstract}
%Active particles often display a strikingly different behaviour with respect to particles in thermal equilibrium. 
We derive an analytic expression for the distribution of velocities of multiple interacting active particles
which we test by numerical simulations.
%This is based on a multidimensional approximation of the Gaussian colored noise driven dynamics and we test it by numerical simulations of the model. 
In clear contrast with equilibrium we find that the velocities are coupled to positions. 
Our model shows that, even for two particles only, the individual velocities display a variance depending on the interparticle separation and the emergence of correlations between the velocities of the particles. When considering systems composed of many particles we find an analytic expression connecting the overall velocity variance to density, at the mean-field level, and to the pair distribution function valid in the limit of small noise correlation times. Finally we discuss the intriguing analogies and main differences between our effective free energy functional and the theoretical scenario proposed so far for phase-separating active particles.
%We discuss how these results are surprisingly similar to those found numerically in the ``run and tumble" model which cannot be treated analytically in these situations.

\end{abstract}

\maketitle

%%%%%%%%%%%%%%%%%%%%%%%%%%%%%%%%%%%%%%%%%%%%%%%%%%%%%%%%%%%%%%%%%%%%%%%%%%%%%%%%%%%%%%%%%%%%%
\section{Introduction.}
%%%%%%%%%%%%%%%%%%%%%%%%%%%%%%%%%%%%%%%%%%%%%%%%%%%%%%%%%%%%%%%%%%%%%%%%%%%%%%%%%%%%%%%%%%%%%
%The velocities of the particles of a classical system in thermal equilibrium behave in a particularly simple way. Indeed, even in very dense and/or cold phases, the distribution of the instantaneous velocities is given by the Maxwell-Boltzmann distribution~\cite{StatLect,StatMechBooks}. This celebrated formula may be expressed as 
%$
%P(\dot{\mathbf{x}})~\propto~\exp ( -\beta~\frac{m}{2} \,\dot{\mathbf{x}} \cdot \dot{\mathbf{x}} )
%$, 
%where the $\dot{\mathbf{x}}$ is the vector of the individual components of the particles velocities, $m$ %is the particles' mass and $\beta = (k_B T)^{-1}$ is the inverse temperature. This implies that the velocities of any equilibrium classical system behave in a quite ``boring'' way, i.e. in the same way of an ideal gas, and in practice statistical mechanics has to deal 
%only with the configurational degrees of freedom.
The velocities of particles in any equilibrium classical system behave in a particularly simple way following the Maxwell-Boltzmann distribution~\cite{StatLect,StatMechBooks}. 
However, when a system is driven out of equilibrium, the situation may change dramatically.
In these systems we do not know generally what is the actual distribution of the velocities
and strong deviations from the Maxwell-Boltzmann distribution may be observed.
In granular systems, for example, non-Gaussian velocity distributions
and cross correlations between the fluctuations of kinetic temperature and density have been 
found both in theoretical models~\cite{granular1,puglisi} and experiments~\cite{granular2}.
The non-equilibrium behaviour of particle velocities is also a fundamental issue in a 
completely different type of far from equilibrium systems that is ``active matter"~\cite{ReviewsActive,
Marchetti13}. 
This novel class of systems may be generally considered as composed by biological or synthetic ``particles'' that are capable of converting available energy into different kinds of persistent motion.
This is the case for example of self-propelled bacteria such as \textit{E. coli}, swimming along straight runs interrupted by random reorientations which can be modelled by the ``run and tumble'' (RT) dynamics~\cite{Berg,Schnitzer,CatesEPL,CatesPRL}. Similarly, Janus-type colloids are propelled by chemical reactions and gradually reorient by rotational Brownian motion which is accounted for by the ``active Brownian'' (AB) model~\cite{Janus1,Janus2,boc2,Janus3,Bocquet}. 
%The dynamics of these systems is well reproduced by schematic models such as the ``run and tumble" 
%dynamics~\cite{Berg,Schnitzer,CatesEPL,CatesPRL} and the ``active Brownian" model~\cite{Janus1,Janus2,Janus3,Bocquet} respectively. 
These systems show a ubiquitous tendency to accumulate near repulsive obstacles~\cite{Walls2,Drop,Fily} and often display clustering and/or phase separation even in the absence of attractive interactions~\cite{boc2,ActiveDepletion,Ivo,Farage,marenduzzo,speckMIPS}.
In Ref.~\cite{CatesPRL} a very general mechanism accounting for these phenomena was proposed,
inspired by the  RT dynamics.  
The basic idea is that the velocities of the particles decrease rapidly 
where the local density increases and conversely the density becomes higher where the local velocity of the particles decreases. This feedback mechanism generates dense regions composed by slow particles and may eventually lead to the so-called ``motility induced phase separation''.
%%%%%%%%%%%%%%%%%%%%%%%%%%%%%%%%%%%%%%%%%%%%%%%%%%%%%%%%%%%%%%%%%%%%%%%%%%%%%%%%%%%%%%%%%%
Despite this elegant scenario was proposed some years ago several fundamental questions still remain unanswered: how do particles velocities depend on density? How do velocities depend on the interaction between particles and on the interparticle distance? To tackle these issues from a theoretical point of view, we consider the Gaussian colored noise (GCN) model~\cite{mucnasp,mucnasoft,Capillars,StructSim} which embodies the persistency of the active motion at the simplest level. Supplementing this model with the \textit{multidimensional unified colored noise approximation} (MUCNA)~\cite{HanggiRev,HanggiUcna,Inertia,UcnaD} we derive an explicit expression for the distribution of velocities of  interacting active particles. We show that this distribution captures well the results of numerical simulation of the GCN model. This distribution shows explicitly that the velocities are coupled to positions via the Hessian matrix associated with the interaction potential. For two interacting particles only the individual velocities have a variance that depends on their distance. Moreover, the model predicts correlations between the velocities of the particles. For an active many-body system we derive an analytic expression connecting the overall velocity variance to the density by a mean-field approximation. Moreover we find an expression connecting the velocity variance to the pair distribution function valid in the limit of small persistence time. We conclude by deriving an effective free energy functional and by comparing it to the one proposed in Ref.~\cite{CatesPRL} discussing the motility induced phase separation scenario for our model
at the mean-field level.

%%%%%%%%%%%%%%%%%%%%%%%%%%%%%%%%%%%%%%%%%%%%%%%%%%%%%%%%%%%%%%%%%%%%%%%%%%%%%%%%%%%%%%%%%%%%%
\section{Main result.}
%%%%%%%%%%%%%%%%%%%%%%%%%%%%%%%%%%%%%%%%%%%%%%%%%%%%%%%%%%%%%%%%%%%%%%%%%%%%%%%%%%%%%%%%%%%%%
The GCN model is defined by the set stochastic differential equations:

\begin{equation} \label{lang}
 \dot{\mathbf{x}} = -\nabla \phi+\bm{\eta}  
\end{equation}

\noindent where the position variables 
$\mathbf{x}=(x_1^1,...,x_1^d,...,x_N^1,...,x_N^d)$, of the $N$ particles in a 
$d$-dimensional space, are driven by the velocities $-\nabla \phi$ (generated by the conservative potential $\phi(\mathbf{x})$). The vector $\bm{\eta}$ is a set of Ornstein-Uhlenbeck processes having
$\langle \eta_i^\alpha \rangle=0$ and  $\langle \eta_i^\alpha(t) \eta_j^\beta(s) \rangle 
= \delta_{ij} \delta_{\alpha \beta} \frac{D}{\tau} \, e^{-|t-s|/\tau}$ (where $i,j=1,...,N$
and $\alpha,\beta=1,...,d$). Here $D$ is the diffusion coefficient of the particles in absence of interactions and $\tau$ is the relaxation time of the ``propulsion'' random forces. 
By using the MUCNA in Ref.~\cite{mucnasp} we have found the approximated stationary probability for the GCN 
model:

\begin{equation} \label{mucna}
\mathcal{P}(\mathbf{x}) = \mathcal{Q}^{-1} \exp \left[ -\frac{\phi(\mathbf{x})}{D} 
-\frac{\tau |\nabla \phi (\mathbf{x}) |^2}{2 D} \right]
||\mathbf{I} + \tau \nabla \nabla \phi(\mathbf{x}) ||
\end{equation}

\noindent where $\mathcal{Q}$ is a normalization factor obtained by integrating over $\mathbf{x}$, $\mathbf{I}$ is the identity matrix, $\nabla \nabla\phi$ is the Hessian of $\phi$ and $||...||$ represents the absolute value of the determinant of a matrix. 
%%%%%%%%%%%%%%%%%%%%%%%%%%%%%%%%%%%%%%%%%%%%%%%%%%%%%%%%%%%%%%%%%%%%%%%%%%%%%%%%%%%%%%%%%%
Within the same approximation we have derived also the conditional probability
$\Pi(\dot{\mathbf{x}}|\mathbf{x})$ of having a velocity vector $\dot{\mathbf{x}}=(\dot{x}_1^1,...,\dot{x}_N^d)$ given that the positions are \textit{fixed} by the vector $\mathbf{x}$:

\begin{equation} \label{PI}
\Pi(\dot{\mathbf{x}}|\mathbf{x}) = \mathcal{N}^{-1} 
\exp \left\{ -\frac{\tau}{2 D} \, \dot{\mathbf{x}} \cdot 
[ \mathbf{I} + \tau \nabla \nabla \phi (\mathbf{x}) ] \cdot \dot{\mathbf{x}} 
\right\}
\end{equation}

\noindent where $\mathcal{N}$ is a normalization factor, obtained by integrating over $\dot{\mathbf{x}}$,
which depends parametrically on the $\mathbf{x}$.
To understand the basic principle of our derivation 
leading to Eq.~(\ref{PI}) we briefly outline it here for a single degree of freedom.
We proceed by further differentiating Eq.~(\ref{lang}) with respect to time and rewrite it as a second-order differential equation containing a white noise term. 
When the dynamics is cast in this form we can obtain a Kramers equation for the full phase-space distribution $\Psi (x,\dot{x};t)$:

\begin{eqnarray} 
& & \frac{\partial \Psi }
{\partial t}
+ \dot{x} \frac{\partial \Psi}{\partial x}
-\frac{1}{\tau} \frac{\partial \phi}{\partial x}
 \frac{\partial \Psi}{\partial \dot{x}} = 
 \nonumber \\
& & \frac{1}{\tau}\frac{\partial}{\partial \dot{x}}
 \left[
 \frac{D}{\tau}
 \frac{\partial}{\partial \dot{x}}
 +\dot{x}\left( 1+\tau 
\frac{\partial^2 \phi}{\partial x^2} 
 \right)
 \right] \Psi
 \label{full1}
\end{eqnarray}

\noindent We now consider the stationary solution
$\Psi_0(x,\dot{x})$. In this case, by multiplying Eq.~(\ref{full1}) by $\dot{x}$ and integrating, we obtain:

\begin{eqnarray}
& & \frac{\partial}{\partial x}
\int d\dot{x} \, {\dot{x}}^2 \Psi_0
+ \frac{1}{\tau}\frac{\partial \phi}{\partial x}
\int d\dot{x} \, \Psi_0
= \nonumber \\
& & 
-\frac{1}{\tau}\left(
1+\tau 
\frac{\partial^2 \phi}{\partial x^2} 
\right)
\int  d\dot{x} \, \dot{x} \Psi_0
 \label{full2}
\end{eqnarray}

\noindent At this point we make the \textit{antsatz} $\Psi_0 = \Pi(x|\dot{x})P(x)$,
where $\Pi$ has the Gaussian form of Eq.~(\ref{PI}), and we substitute this into Eq.~(\ref{full2}). 
With such a factorization Eq.~(\ref{full2}) reduces to an ordinary differential equation for $P(x)$, whose solution 
 coincides precisely with the probability $\mathcal{P}$ of
Eq.~(\ref{mucna}) proving that $\Pi$ is the correct velocity distribution 
within our approximation (see also~\cite{supp} for details
on the derivation for many degrees of freedom).

\begin{figure}
\begin{center}
\includegraphics[width=0.35\textwidth]{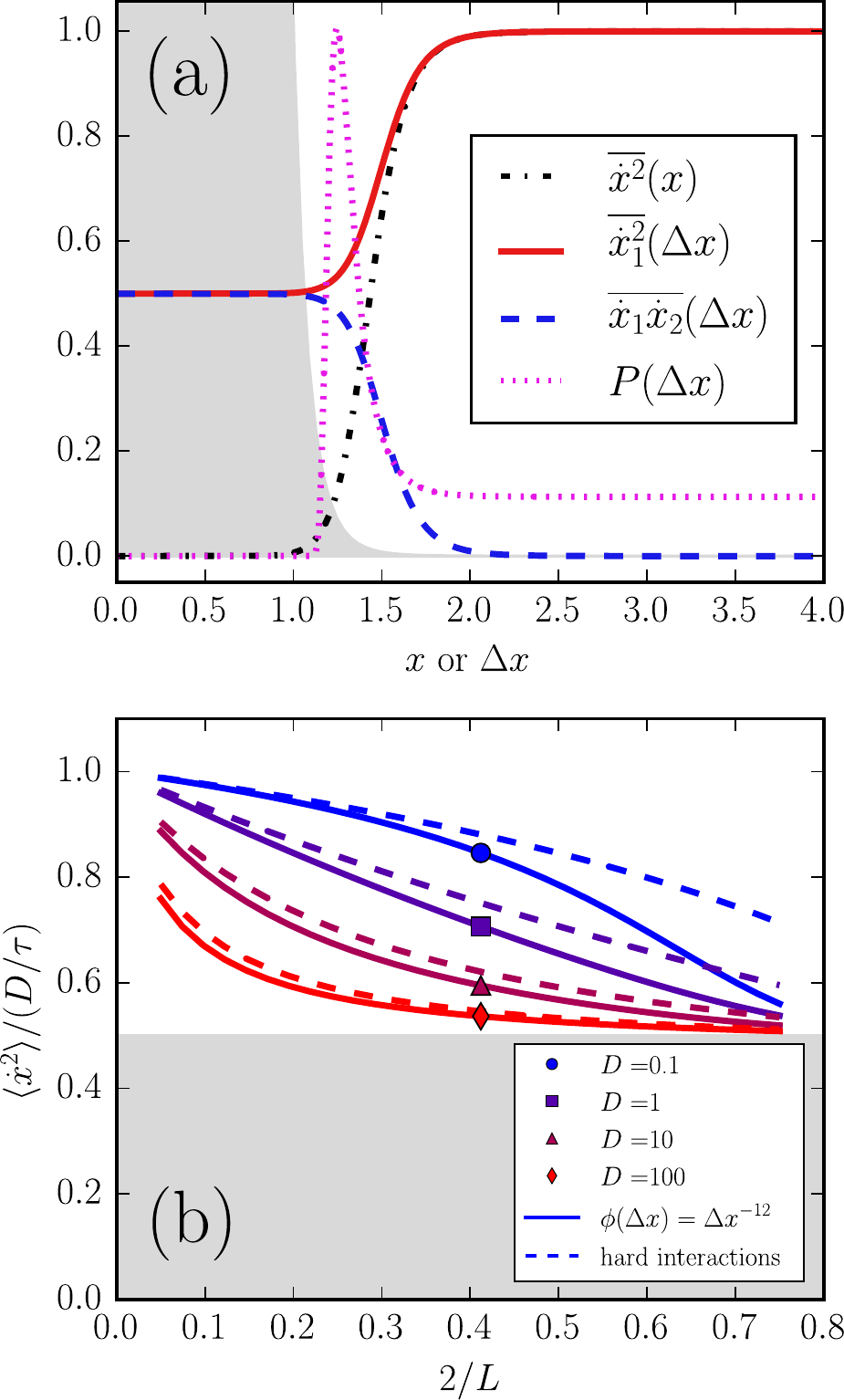}
\caption{ 
(\textbf{a}) Velocity variance and correlation as a function of distance for $D=1$ and $\tau=1$
as computed from Eq.~(\ref{PI}). 
The dashed-dotted line is the variance of velocity as a function of $x$ 
for one particle moving in 1$d$ in the presence of the repulsive barrier $\phi = x^{-12}$ (shaded area).
The full line is the variance of velocity of one particle interacting with another particle via the repulsive potential $\phi = \Delta x^{-12}$ (shaded area), the dashed line is the velocity correlation between the two interacting particles. The dotted line is the probability 
distribution given by Eq.~(\ref{mucna}) for the interacting particles and $L=8$.
(\textbf{b}) Overall velocity variance (Eq.~(\ref{v2})) 
for two interacting active particles as a function of density for fixed $\tau=1$. 
Full lines with one symbol represent $\langle \dot{x}^2 \rangle$ for the interaction $\phi=\Delta x^{-12}$ at different values of $D$ (see legend). The dashed line $\langle \dot{x}^2 \rangle$ in the limiting case of two hard spheres. The shaded area represents the lower bound $D/(2 \tau)$.
}
\label{fig:f1}
\end{center}
\end{figure}

The result of Eq.~(\ref{PI}) immediately tell us that the probability distribution of $\dot{\mathbf{x}}$ is a multivariate Gaussian but, very differently from the Maxwell-Boltzmann distribution, its covariance matrix depends on the positions via the Hessian.
%%%%%%%%%%%%%%%%%%%%%%%%%%%%%%%%%%%%%%%%%%%%%%%%%%%%%%%%%%%%%%%%%%%%%%%%%%%%%%%%%%%%%%%%%%
Let us start examining the implications of Eq.~(\ref{PI}) by considering a single active particle moving in one dimension (1$d$) and subjected to an external potential. In this case Eq.~(\ref{PI}) gives the variance as a function of $x$: 
$\overline{{\dot{x}^2}}(x)= \int_{-\infty}^{\infty} d\dot{x} \, \Pi(\dot{x}|x) \, \dot{x}^2  =(D/\tau)/[1+\tau \phi^{\prime \prime}(x)]$, where we use the overbar to indicate the averaging over $\dot{x}$ and the prime represents the derivative with respect to $x$. This shows that when the particle is in a region of high potential curvature its velocity variance decreases. To be more specific let us consider a purely repulsive potential of the form $\phi=x^{-12}$. The corresponding $\overline{{\dot{x}^2}}$ is plotted in Fig.~\ref{fig:f1}(a) as a dashed-dotted line and this shows clearly that the velocity variance is close to the unperturbed value $D/\tau$ where the $\phi^{\prime \prime}$ is small, but it decreases rapidly to zero in proximity to the repulsive ``wall'' generated by the external potential.

%%%%%%%%%%%%%%%%%%%%%%%%%%%%%%%%%%%%%%%%%%%%%%%%%%%%%%%%%%%%%%%%%%%%%%%%%%%%%%%%%%%%%%%%%%%%%
\section{Two interacting particles.}
%%%%%%%%%%%%%%%%%%%%%%%%%%%%%%%%%%%%%%%%%%%%%%%%%%%%%%%%%%%%%%%%%%%%%%%%%%%%%%%%%%%%%%%%%%%%%
We now consider two particles, with positions $\mathbf{x}=(x_1,x_2)$, interacting via the potential 
$\phi(x_1-x_2)=\phi(\Delta x)$ and moving in 1$d$. In this case Eq.~(\ref{PI}) can be used to compute both the velocity variance: $\overline{{\dot{x}_1}^2}(\Delta x)=\overline{{\dot{x}_2}^2}(\Delta x)
=(D/\tau) [1+\tau \phi^{\prime \prime}(\Delta x)]/[1+2 \tau \phi^{\prime \prime}(\Delta x)]$
and the velocity correlations: 
$\overline{\dot{x}_1 \dot{x}_2}(\Delta x)=D \phi^{\prime \prime}/[1+2 \tau \phi^{\prime \prime}(\Delta x)]$.
These equations show that the variance of the velocity decreases when the particles are at a distance where the interaction potential has high curvature and that in this situation also a correlation of velocities emerge. To visualize these quantities we consider the interaction potential $\phi(\Delta x)= {\Delta x}^{-12}$ and we plot $\overline{{\dot{x}_1}^2}$ and $\overline{\dot{x}_1 \dot{x}_2}$ in Fig.~\ref{fig:f1}(a) as a full and dashed line respectively. Since this potential has a curvature that goes rapidly to zero, 
$\overline{{\dot{x}_1}^2}$ tends rapidly to the unperturbed value $D/\tau$, while when the particles are close enough the value of $\overline{{\dot{x}_1}^2}$ goes to $D/(2\tau)$. This limiting value represents the mean squared speed of the center of mass of the two particles system. 
Similarly, the correlation $\overline{\dot{x}_1 \dot{x}_2}$ goes to zero at large $\Delta x$ (where interaction is small) while, when $\Delta x$ is small, the two particles will move coherently with the velocity of the center of mass resulting in the limiting value  $\overline{\dot{x}_1 \dot{x}_2} = D/(2\tau)$.
%%%%%%%%%%%%%%%%%%%%%%%%%%%%%%%%%%%%%%%%%%%%%%%%%%%%%%%%%%%%%%%%%%%%%%%%%%%%%%%%%%%%%%%%%%

Up to this point we have considered the velocity variance when the particles positions are fixed arbitrarily. 
However, we want also to compute this quantity averaging it over the positions to obtain the overall velocity variance of a GCN-driven system.
This can be done, within the MUCNA, by combining Eq.~(\ref{mucna}) and~(\ref{PI}): 

\begin{equation} \label{v2}
\langle \dot{x}^2 \rangle = \frac{1}{d N} \int d\mathbf{x} \, \mathcal{P}(\mathbf{x}) 
\int d\dot{\mathbf{x}} \; \dot{\mathbf{x}} \cdot \dot{\mathbf{x}} \; \Pi(\dot{\mathbf{x}}|\mathbf{x}) 
\end{equation}

\noindent where, by dividing by $d N$, we average also over all particles and all components. 
To understand this result let us consider, as above, two particles in $1d$ interacting via 
$\phi=\Delta x^{-12}$ and assume that they move in a box of length $L$ with periodic boundary conditions. 
The $\langle \dot{x}^2 \rangle$ computed numerically via Eq.~(\ref{v2}) is plotted in Fig.~\ref{fig:f1}(b) (full lines) as a function of the 1$d$ density $\rho=2/L$ of the system and for several values of $D$ (at fixed $\tau$). 
In Fig.~\ref{fig:f1}(b) we divide the variance by the free-particle value $D/\tau$ so that 
$\langle \dot{x}^2 \rangle/(D/\tau)$ reduces to unity in absence of interactions.
This shows clearly that $\langle \dot{x}^2 \rangle/(D/\tau)$ decreases systematically with increasing $\rho$ and with increasing $D$. Upon increasing $D$ we also observe a change in the convexity of the curves in   Fig.~\ref{fig:f1}(b). Such a phenomenology can be understood by noting that the $\mathcal{P}(\Delta x)$ is very peaked around $|\Delta x| \approx 1$ (see dotted line in Fig.~\ref{fig:f1}(a)). Note also that the $\mathcal{P}(\Delta x)$ is practically zero when $|\Delta x| < 1$ defining the ``diameter'' of the particles $\sigma \approx 1$ that changes very little in the wide range of $D$ and $\tau$ studied here. 
In Ref.~\cite{mucnasp} we have shown that, by assimilating the potential $x^{-12}$ to a hard potential, $\mathcal{P}(\Delta x)$ can be approximated by a Dirac delta of area $\mathcal{Q}^{-1}\sqrt{2 D \tau}$ around the points $\Delta x \approx \pm \sigma$  and reduced to $\mathcal{Q}^{-1}$ elsewhere, the normalization constant being $\mathcal{Q} = (L-2\sigma)+2\sqrt{2 D \tau}$. As discussed above in this limit we find also: $\overline{{\dot{x}_1}^2} = D/\tau$ if $|\Delta x|>\sigma$, and $\overline{{\dot{x}_1}^2} = D/(2 \tau)$ if 
$|\Delta x| \leq \sigma$. Combining these formulae we find the velocity variance 
of two GCN active hard spheres: 
$\langle \dot{x}^2 \rangle/(D/\tau) = \frac{2 \rho^{-1}-2 \sigma+\sqrt{2 D \tau}}{2 \rho^{-1}-2 \sigma+ 2\sqrt{2 D \tau}}$ which is a monotonic decreasing function of $\rho$ and $D$. 
Moreover $\langle \dot{x}^2 \rangle /(D/\tau)$ is bounded from below by $1/2$ (in the limit of both large 
$\rho$ and $D$). This is plotted as a dashed lines in Fig.~\ref{fig:f1}(b) and follows qualitatively well the results of the $\phi=\Delta x^{-12}$ case. Moreover from this equation we can see that $\partial^2 \langle \dot{x}^2 \rangle/\partial \rho^2 > 0$ if $\mathcal{L}=\sqrt{D \tau}>\sigma/\sqrt{2}$ where we have intorduced the characteristic length of the active motion $\mathcal{L}$. This, in practice, means that, the velocity variance becomes especially sensitive to density changes when the \textit{P\'{e}clet number}~\cite{marenduzzo} $Pe$ becomes of order one, i.e. $Pe = \mathcal{L}/\sigma \gtrsim 1$.

\begin{figure*}
\begin{center}
\includegraphics[width=0.95\textwidth]{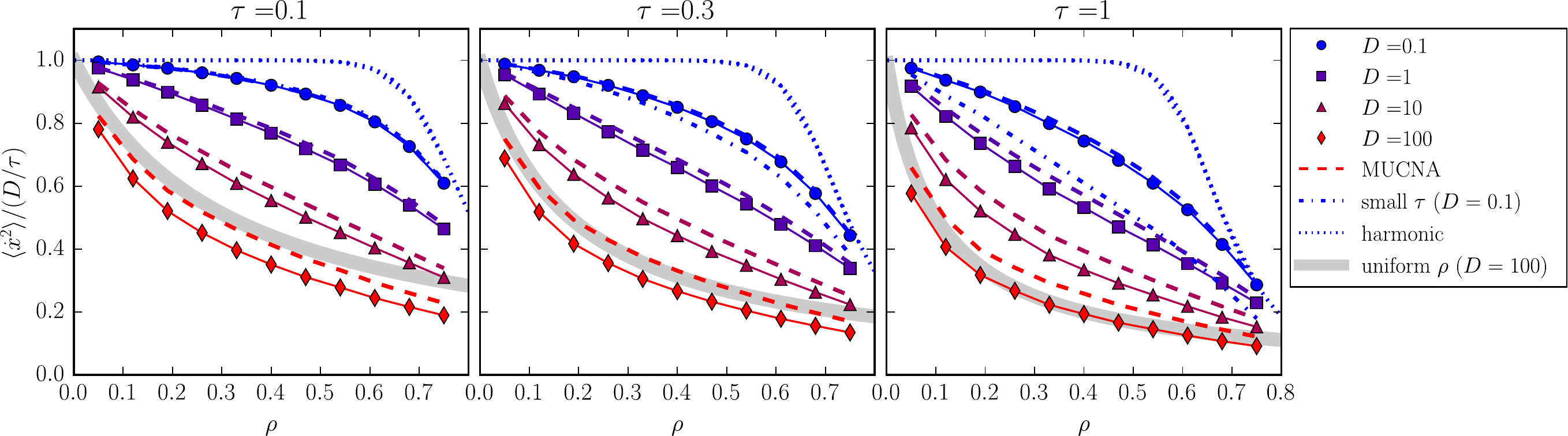}
\caption{Normalized velocity variance for a $1d$ system of many interacting active particles. Symbols are the results of numerical simulations for several values of $\tau$ and $D$ (see legend). 
Dashed lines are the theoretical velocity variances obtained by averaging Eq.~(\ref{PI})  over the coordinates obtained numerically. Thick lines are the result of a homogenous density approximation.
Dashed-dotted lines represent the small-$\tau$ approximation connecting the variance to the pair distribution function. Dotted lines are the velocity variances obtained by mapping the system onto a harmonic model.
}
\label{fig:f2}
\end{center}
\end{figure*}

%%%%%%%%%%%%%%%%%%%%%%%%%%%%%%%%%%%%%%%%%%%%%%%%%%%%%%%%%%%%%%%%%%%%%%%%%%%%%%%%%%%%%%%%%%%%%
\section{Many interacting particles.}
%%%%%%%%%%%%%%%%%%%%%%%%%%%%%%%%%%%%%%%%%%%%%%%%%%%%%%%%%%%%%%%%%%%%%%%%%%%%%%%%%%%%%%%%%%%%%
To make progress towards the statistical description of the velocities of the many-body active system we 
consider $N$ interacting particles in 1$d$. We perform numerical simulations of  systems with $N=1000$ composed by GCN-driven particles interacting via the potential $\phi(\mathbf{x})=\sum_{i>j} (x_i-x_j)^{-12}$ for several  values of the density $\rho=N/L$, of $D$ and $\tau$. In all these simulations we compute the variance $\langle \dot{x}^2 \rangle$ and report the results in Fig.~\ref{fig:f2} as connected symbols. Qualitatively the emerging scenario seems to be close to the two-particle case (see Fig.~\ref{fig:f1}(b)). However from a quantitative point of view the two-particle model is far from the results in Fig.~\ref{fig:f2}, in particular the many-body $\langle \dot{x}^2 \rangle/(D/\tau)$ reaches values  well below the lower bound of the two-particle case $1/2$.
%%%%%%%%%%%%%%%%%%%%%%%%%%%%%%%%%%%%%%%%%%%%%%%%%%%%%%%%%%%%%%%%%%%%%%%%%%%%%%%%%%%%%%%%%%
To test uniquely the approximated distribution given by Eq.~(\ref{PI}), we compute the average 
over positions in Eq.~(\ref{v2}) directly from the coordinates obtained numerically, instead of using the 
theoretical $\mathcal{P}$ of Eq.~(\ref{mucna}).
This is plotted in Fig.~\ref{fig:f2} as dashed lines and follows well the numerical curves, although some expected deviation~\cite{HanggiRev} is observed upon increasing $D$ to very high values.

\begin{figure}
\begin{center}
\includegraphics[width=0.3\textwidth]{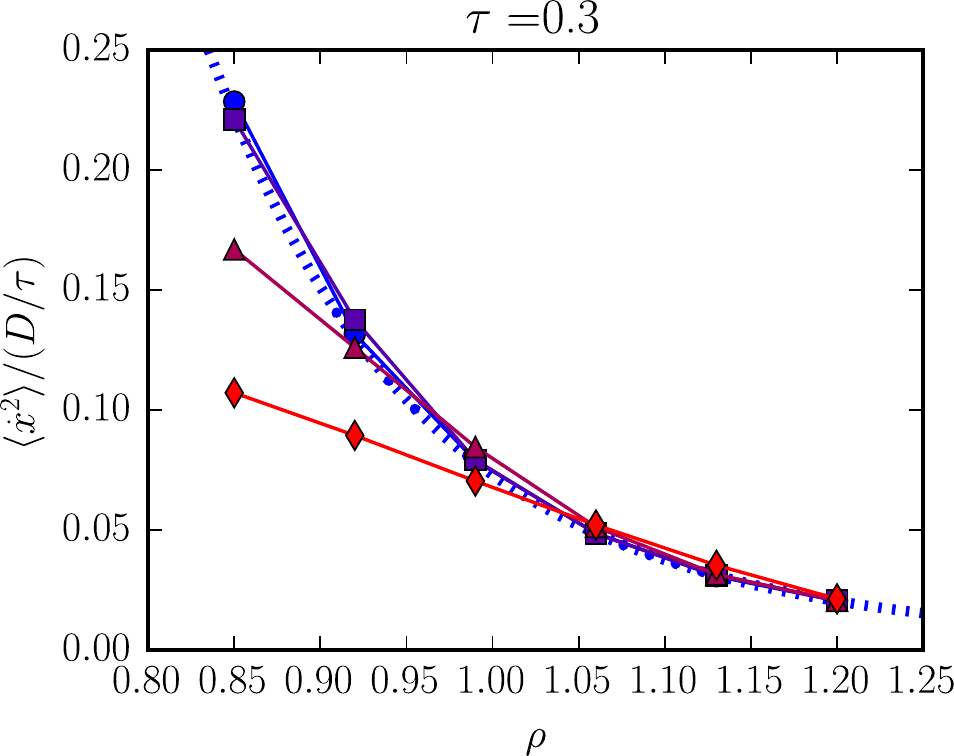}
\caption{ Comparison between the harmonic model (dotted line) and the numerical simulations 
(symbols) at high $\rho$ for several values of $D$ (same legend as Fig.~\ref{fig:f2}) 
at fixed $\tau=0.3$.
}
\label{fig:fh}
\end{center}
\end{figure}

If we assume a uniform density and long-ranged interactions (mean-field approximation) the velocity distribution (Eq.~(\ref{PI})) simplifies substantially since all the out-of-diagonal term of 
$\nabla \nabla \phi$ are of order one and can be neglected with respect to the terms on the main diagonal that are of order $N$~\cite{mucnasoft}. This yields the density-dependent variance:

\begin{equation}\label{MF}
\frac{\langle \dot{x}^2 (\rho) \rangle} 
{(D/\tau)} = \frac{1}{1+\tau \phi_2 \rho} = \frac{1}{1+\rho \mathcal{L}}
\end{equation}
 
\noindent where $\phi_2 = \int_\sigma^\infty dx \, \phi^{\prime \prime}(x)$ is the mean potential 
curvature integrated up to the diameter $\sigma$.
In the last equality of Eq.~(\ref{MF}) we have used
the fact that, for a generic repulsive potential, 
$\sigma$ corresponds roughly to the distance where the interaction force balances the self-propulsion 
(i.e. $|\phi^\prime(\sigma)| \approx \sqrt{D/\tau}$) and 
$\phi_2 = \phi^\prime (\infty)-\phi^\prime (\sigma) \approx \sqrt{D/\tau}$.
This is plotted in Fig.~\ref{fig:f2} as a thick line for the largest $D$ 
and follows well the data when $\mathcal{L}$
is large.
%, this is because in the large $D$ and $\tau$ regime a large fraction of particles are found in contact and strongly interacting. 
%In this situation the curvature experienced by the particles is close to the mean curvature $\phi_2$.
%%%%%%%%%%%%%%%%%%%%%%%%%%%%%%%%%%%%%%%%%
When $\tau$ is small Eq.s~(\ref{mucna}) and (\ref{PI}) can be expanded to first order in $\tau$ (see~\cite{supp}) which gives an alternative formula for $\phi_2$ in terms of the pair distribution function $g(x)$, i.e. $\phi_2 = \int_0^\infty dx \, g(x) \phi^{\prime\prime}(x) $. This is plotted in Fig.~\ref{fig:f2} as dashed-dotted lines and compares well with the numerical simulations at small values of $\tau$. However, by fixing $\tau$ and increasing $D$ this approximation deviates strongly from the simulations 
and therefore it is not shown in Fig.~\ref{fig:f2}.
%%%%%%%%%%%%%%%%%%%%%%%%%%%%%%%%%%%%%%%%%%%%%%%%%%%%%%%%%%%%%%%%%%%%%%%%%%%%%%%%%%%%%%%%%%
In order to describe more accurately the high-density regime, we derive also a harmonic model for the velocity distribution. To this aim we consider a $1d$ system of active particles connected by springs having elastic constant $k$. In this system each particle is connected only to its nearest neighbour by 
a harmonic potential, the Hessian matrix in Eq.~(\ref{PI}) does not depend on the positions anymore and it takes the form of a banded symmetric Toeplitz matrix whose elements are $[\mathbf{I}+\tau \nabla \nabla \phi(\mathbf{x})]_{ii} = 1+ 2 k \tau $, $[\mathbf{I}+\tau \nabla \nabla \phi(\mathbf{x})]_{i \, i\pm1}=-k \tau$ and zero elsewhere. The eigenvalues of this matrix are known~\cite{matrix} and the mean of their inverse can be computed as a Watson integral giving the result:
$\langle {\dot{x}}^2 \rangle/(D/\tau) = [(1+2k\tau)^2-4 k^2\tau^2]^{-1/2}$. To compare this result with the simulations of the $\phi={\Delta x}^{-12}$ potential we assume that in the high $\rho$ regime particles are separated by the average distance $\rho^{-1}$ and expand the interaction potential to 2nd order around these distance.  We obtain an estimate of the effective spring constant as $k(\rho)=156 \, \rho^{14}$. The resulting $\langle {\dot{x}}^2 \rangle/(D/\tau)$ is plotted in Fig.~\ref{fig:f2} as a dotted line and it is found to well reproduce  the numerical results at high values of $\rho$ and $\tau$, while some deviation is observed upon increasing $D$. However, by going at high densities ($\rho \approx 1$) this harmonic model predicts quantitatively very well $\langle {\dot{x}}^2 \rangle/(D/\tau)$ for all values of $D$ and $\tau$ (see Fig.~\ref{fig:fh}).

%%%%%%%%%%%%%%%%%%%%%%%%%%%%%%%%%%%%%%%%%%%%%%%%%%%%%%%%%%%%%%%%%%%%%%%%%%%%%%%%%%%%%%%%%%%%%
\section{Effective free energy functional.}
\begin{figure}
\begin{center}
\includegraphics[width=0.5\textwidth]{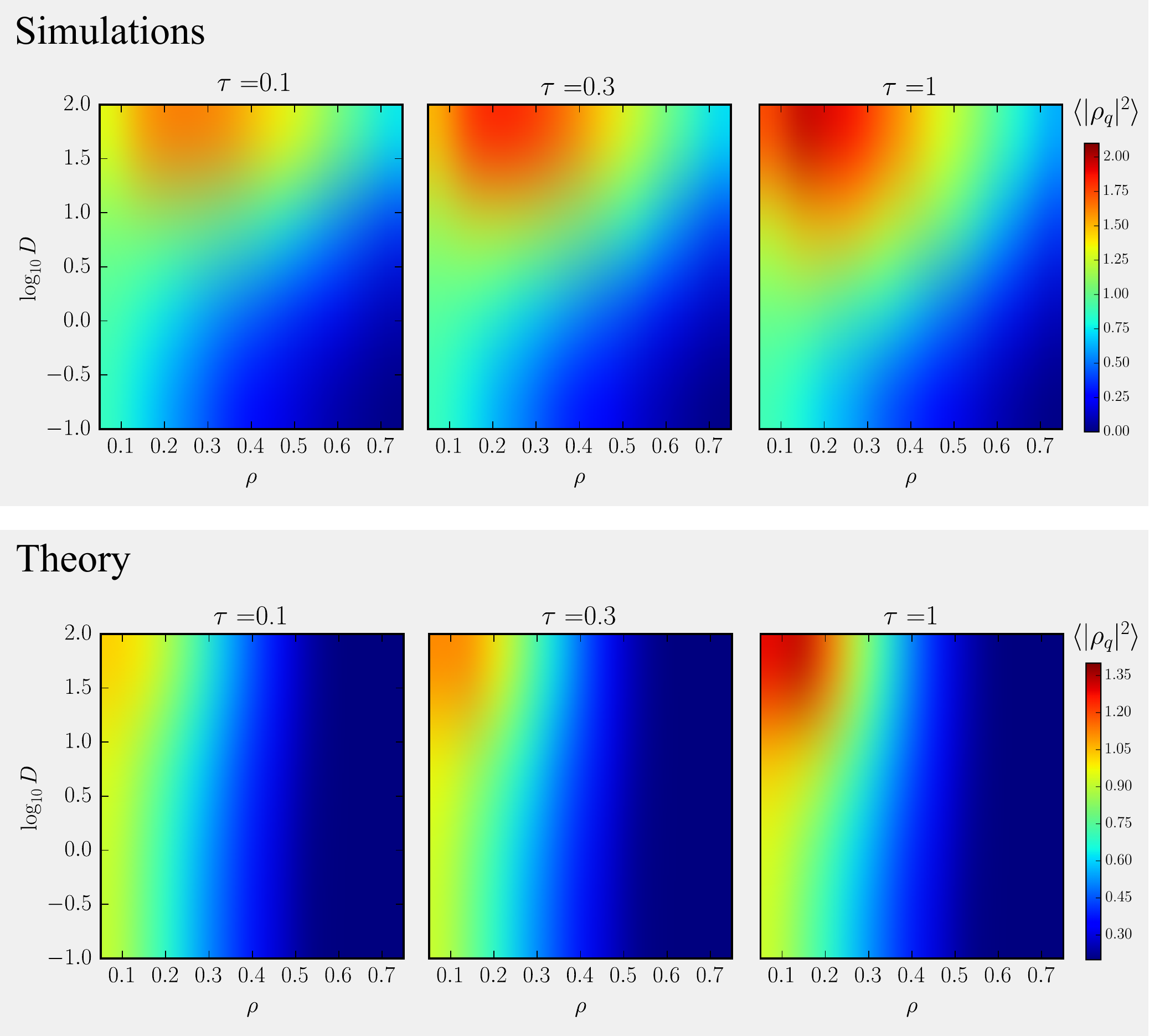}
\caption{ (Top-panel) The colormap represents 
 $\langle |\rho_q|^2 \rangle$  at low $q$ obtained numerically
for several values of $\rho$, $D$ and $\tau$. Colouring close to red indicates
large density fluctuations at that state-point $(\rho,D)$.
(Bottom-panel) Theoretical value of $\langle |\rho_q|^2 \rangle$ (computed from Eq.~(\ref{FEMF}))
shows a qualitative behaviour very similar to the one found in simulations.
}
\label{fig:f3}
\end{center}
\end{figure}

%%%%%%%%%%%%%%%%%%%%%%%%%%%%%%%%%%%%%%%%%%%%%%%%%%%%%%%%%%%%%%%%%%%%%%%%%%%%%%%%%%%%%%%%%%%%%
We have seen that the  probability distribution given by Eq.~(\ref{mucna}) can be mapped onto a Boltzmann distribution characterized by the effective potential 
${\mathcal{H}=\phi+\tau |\nabla \phi |^2/2 
-D \ln||\mathbf{I} + \tau \nabla \nabla \phi ||}$.
Exploiting this analogy, as we have shown in Ref.~\cite{mucnasoft}, 
we can construct an effective free energy functional
of the form 
${\mathcal{F}[\mathcal{P}]=\int d\mathbf{x}\,\mathcal{P}(\ln\mathcal{P}+\mathcal{H}/D)}$.
Moreover by using Eq.~(\ref{PI}) we can rewrite 
${\ln ||\mathbf{I} + \tau \nabla \nabla \phi|| = -\ln \left| \left| \tau \, \overline{\dot{x}_i^{\alpha} \, \dot{x}_j^{\beta}}/D \right| \right|}$ since the determinant of the inverse is the inverse of the determinant. Using these results we 
can recast $\mathcal{F}$ as 

\begin{equation} \label{FE}
\frac{\mathcal{F}[\mathcal{P}]}{D}=\int d\mathbf{x}\,\mathcal{P} \left( \ln\mathcal{P}
+ \ln \left| \left| \frac{\tau}{D} 
\, \overline{\dot{x}_i^{\alpha} \, \dot{x}_j^{\beta}} \right| \right|+
\frac{\phi}{D}+\frac{\tau}{2 D} |\nabla \phi |^2 \right)
\end{equation}

\noindent which shows a dependence on the velocities resembling closely the one suggested 
in Ref.~\cite{CatesPRL}. In particular the term $\ln \left| \left| \, \overline{\dot{x}_i^{\alpha} \, \dot{x}_j^{\beta}} \right| \right|$ represents the differential entropy of the velocity distribution,
i.e. how much velocities are ``spread'' in the velocity space. This implies that if velocities have some very high values, for a given configuration $\mathbf{x}$, this brings positive contribution 
to free energy and therefore $\mathcal{P}$ will be low for such $\mathbf{x}$. 
Conversely, if velocities concentrate around zero at $\mathbf{x}$, $\mathcal{P}$ will grow for this configuration. 
%%%%%%%%%%%%%%%%%%%%%%%%%%%%%%%%%%%%%%%%%%%%%%%%%%%%%%%%%%%%%%%%%%%%%%%%%%%%%%%%%%%%%%%%%%%%%%%%%%%%%%
Note also that we know how to express the velocity term in Eq.~(\ref{FE}) as a function of density, at least at the mean field level in $1d$, by using Eq.~(\ref{MF}). 
This further suggests that the velocity variance is itself a decreasing function of density
establishing the feedback mechanism 	hypothesized in Ref.~\cite{CatesPRL}.
Moreover, if we assume a homogenous $\rho$ and only repulsive interactions, the terms $\phi$ and $|\nabla \phi|^2$ correspond only to repulsive potential terms in Eq.~(\ref{FE}).
With these considerations we rewrite Eq.~(\ref{FE}) in the mean-field form:

\begin{equation} \label{FEMF}
f(\rho) = \rho \left[ 
\ln \left( \frac{\rho}{1-\rho\sigma} \right)-1 \right] 
- \int_0^\rho d \rho^\prime \ln ( 1+\rho^\prime \mathcal{L})
\end{equation}

\noindent where we have introduced the free energy per unit length $f=\mathcal{F}/L$ and we have absorbed all the repulsive terms in the $1d$ hard-spheres excess free energy $-\rho \ln (1-\rho\sigma )$.
If $\partial^2_{\rho^2} f <0$ the homogenous density phase is unstable and the system undergoes a spinodal decomposition. In contrast to this scenario, we find from the mean-field Eq.~(\ref{FEMF}) always gives $\partial^2_{\rho^2} f >0$. This suggests that our active system does not phase separate for any value of $\rho$, $D$ and $\tau$ even in the presence of long-ranged interactions. This is consistent with the numerical results that do not show any discontinuity in any of the average values that we have monitored. However Eq.~(\ref{FEMF}) does predict an anomalous behaviour of density fluctuations. To show this we consider the $1d$ Fourier transformed density fluctuations $\langle |\rho_q|^2 \rangle$, where $\rho_q = N^{-1/2} \sum_i e^{i \,  q \, x_i}$ ($q$ being the wavevector). We study  the  long wave-length  density fluctuations by choosing
$q \approx {(20 \sigma)^{-1}}$ and compute  $\langle |\rho_q|^2 \rangle$ in numerical simulations which is plotted in Fig.~\ref{fig:f3}~(top panel) as a colormap for several values of $D$, $\rho$ and $\tau$.
This shows that at sufficiently high $D$ the $\langle |\rho_q|^2 \rangle$ develops a maximum as a function of the average density, $\rho$, and that this maximum increases in amplitude as $\tau$ increases.
At such small $q$ we can approximate~\cite{vank} $\langle |\rho_q|^2 \rangle \approx |\rho \partial^2_{\rho^2}f|^{-1}$ which is plotted in Fig.~\ref{fig:f3}~(bottom panel). 
This is found to follow qualitatively very well the numerical results in the whole $D$, $\tau$ and $\rho$ range explored. In practice, the effect of the colored noise is to introduce a sort of weak effective attraction at high persistence lengths $\mathcal{L}$ enhancing the density fluctuations in the low-density regime with respect to a purely repulsive equilibrium system. In this low-$\rho$ /high-$\mathcal{L}$ regime an evident clustering of the particles is generated by the persistent propulsion forces.

%%%%%%%%%%%%%%%%%%%%%%%%%%%%%%%%%%%%%%%%%%%%%%%%%%%%%%%%%%%%%%%%%%%%%%%%%%%%%%%%%%%%%%%%%%%%%
\section{Conclusions.}
%%%%%%%%%%%%%%%%%%%%%%%%%%%%%%%%%%%%%%%%%%%%%%%%%%%%%%%%%%%%%%%%%%%%%%%%%%%%%%%%%%%%%%%%%%%%%
By using the unified colored noise approximation, we have derived an explicit expression for the distribution of velocities of interacting active particles and shown that it agrees well with the numerical simulation of the GCN model. This distribution predicts the non-equilibrium coupling between velocities and positions and the correlation between velocities of different particles. These have been characterized in detail for the two-particle case. In the case of many interacting active particles, we have derived approximations connecting the velocity variance to the pair distribution function at low values of the persistence time. From our microscopic model, we have also derived directly an effective free energy functional that shows surprising similarities with the one proposed so far for describing the phase separation in active particles. Our functional establishes directly a connection between the stationary probability of coordinates and the covariance matrix of velocities showing that a configuration with low velocities is favoured. Moreover, our theory shows that when interactions are strong (i.e. when density is high) the velocities tend to decrease generating a feedback mechanism that enhances density fluctuations.
The basic physics of this phenomenon is captured by a mean-field version of our functional that predict qualitatively well the excess density fluctuations observed numerically.
These anomalous density fluctuations indicate clustering of the particles in analogy with the cluster formation observed experimentally for Janus colloids~\cite{boc2}.
However, in the present one-dimensional  mean-field model, this mechanism is too weak to lead to spinodal decomposition.
This seems in contradiction with the results of Ref.~\cite{Farage} in which a further approximation of the same same theoretical framework has been found to predict very well the phase separation of AB particles. With this perspective it would be interesting to extend our mean-field model to the $3d$ case, accounting explicitly for the potential terms, and perform exhaustive $3d$ simulations of the GCN model.

The research leading to these results has received funding from
the European Research Council under the European Union's
Seventh Framework Programme (FP7/2007-2013)/ERC grant
agreement no. 307940. NG acknowledges support from MIUR
(``Futuro  in  Ricerca"  ANISOFT/RBFR125H0M).
MP acknowledges support from NSF-DMR-1305184
%We also acknowledge NVIDIA for hardware donation.

\end{document}